\documentclass[11pt,twoside]{article}

\usepackage{asp2006}
\usepackage{epsf}
\usepackage{psfig}
\usepackage{lscape}

\begin{document}
\title{SUPERSOFT X-RAY LIGHT CURVE OF RS OPH \\-- THE WHITE DWARF MASS IS NOW INCREASING}   %%% Fill in title

\begin{quote}
Mariko Kato,$^1$ Izumi Hachisu,$^2$ Gerardo Juan Manuel Luna$^3$\\
\vspace{-0.1cm}\\
%\indent\vdots\\
\footnotesize
{\itshape $^1$Keio University, Yokohama 223-8521, Japan}\\
%\vspace{-0.1cm}\\
{\itshape $^2$Univ. of Tokyo, Tokyo 153-8902, Japan}\\
%\vspace{-0.1cm}\\
%\vspace{-0.1cm}\\
{\itshape $^3$University of Sa\o Paulo, 05508-900 Sao Paulo, Brazil}\\
%\vspace{-0.1cm}\\
\normalsize
\end{quote}

\begin{abstract} 
The recurrent nova RS Ophiuchi, one of the candidates for Type Ia 
supernova progenitors, underwent the sixth recorded outburst
in February 2006. We report a complete light curve of supersoft X-ray 
that is obtained for the first time. A numerical table of X-ray data is provided. 
The supersoft X-ray flux emerges about 30 days after the optical peak 
and continues until about 85 days when the optical flux shows the final decline. 
Such a long duration of supersoft X-ray phase can be naturally 
understood by our model in which a significant
amount of helium layer piles up beneath the hydrogen burning zone
during the outburst, suggesting that the white dwarf mass is effectively
growing up.  We have estimated the white dwarf mass in RS Oph
to be $1.35 \pm 0.01~ M_\odot$ and its growth rate 
to be  about $(0.5-1) \times 10^{-7} M_\odot$ yr$^{-1}$ in average.
\end{abstract}

\section{Optical and Supersoft X-ray Light Curves of RS Oph}

RS Oph is one of the well-observed recurrent novae and is suggested to be a progenitor
of Type Ia supernova. It has undergone its sixth recorded outburst on 2006 February 12 \citep{nar06}
and many observational results are reported (see other papers in this proceedings).
It's $y$ magnitude light curve has been obtained throughout the outburst \citep{hac06}
[the numerical table is provided in \citet{hac08} in this volume], 
which shows a mid-plateau phase that lasts 45-75 
days from the optical peak followed by a quick decrease (Figure 1). 

RS Oph has also been observed with X-ray satellites. 
We analyzed the {\it Swift} XRT observations available in the HEAS-ARC' database and 
extracted the count rate in the energy band of 0.3-0.55 keV binned in 2000 s.  
\citep[see][for more details]{hac07}. 
The supersoft X-ray (SSX) light curve is plotted in Figure 1 (see also Table 1).
The light curve rises at about 30 days after the optical peak and shows 
a long plateau phase that lasts as long as about 50 days 
corresponding to a long mid-plateau phase of optical light curve.

\section{Modelling of SSX Phase}

During the nova outburst hydrogen-rich envelope around the white dwarf (WD) 
expands to a giant size and strong wind mass-loss occurs. In such stages  
dynamical calculation codes often encounter numerical difficulties, so we 
cannot calculate light curves. 
For example, one must take off the outermost Lagrange mesh points, which prevents
us accurately determining the wind mass-loss rate and calculating the resultant 
evolution speed of novae.

We have calculated light curve models based on the optically thick wind 
theory \citep{kat94h}, which is a quasi-evolution Euler code in which 
the wind mass-loss rate is accurately obtained as an eigenvalue of a boundary 
value problem. The photospheric temperature and luminosity are also accurately 
calculated. Therefore, up to now, the optically thick wind is only the method 
that can follow the theoretical light curves of novae.

To explain the SSX phase and the optical light 
curves of RS Oph, we have included effects of heat exchange between 
the hydrogen-rich envelope and a helium layer underneath. 
Hydrogen burning produces hot helium ash which accumulates underneath the burning 
zone because convection may descend quickly after the optical peak. This helium 
layer grows in mass with time and behave as a heat reserver.  
In the later phase of the outburst heat flows upward from the hot helium layer, 
which keeps hydrogen-rich envelope hot enough  
to emit SSX in a long time. After calculating many models for two parameters, 
i.e., the WD mass and 
the hydrogen content of the envelope, we obtain a best fit model as shown in Fig 1a  
\citep[see][for details]{hac07}. This model reproduces the optical light curve 
and explains reasonably well the X-ray count rate. 

Figure 1b and Table 2 show that the wind mass-loss stops when the SSX count rate increases.  
The total luminosity $L$ is almost constant until day 80 when nuclear burning 
extinguishes and helium ash layer becomes too cool to provide heat any more. 
In this way the duration of the SSX phase can be explained only if we 
assume hot helium ash underneath the hydrogen layer. This helium layer accumulates 
on the WD although some part of the hydrogen-rich matter is blown off by the wind.
Therefore, we conclude that the WD mass is growing though the 2006 outburst.

\section{Summary}

We summarize our results as follows;

\noindent
1. The WD mass is estimated to be $1.35 \pm 0.01~ M_\odot$ from the 
optical and X-ray light curve fittings.

\noindent
2. The accreted matter during  21 years before the outburst is estimated 
from the envelope mass at the optical peak to be $4 \times 10^{-6}M_{\odot}$,   
$50-70\%$ of which is ejected by the wind and the rest $30-50\%$ accumulates on 
the WD. Therefore the net growth rate of the WD is $(0.5-1) \times 10^{-7}M_{\odot}$yr$^{-1}$.

\noindent
3. The durations of the mid-plateau phase of optical and the peak 
plateau phase of SSX suggest the presence of a helium layer which accumulates 
on the WD. Therefore, the WD mass of RS Oph is now growing. 
 
\noindent
4. The hydrogen content is expected to decrease by $\Delta X = 0.2-0.4$
during the outburst, because convection may quickly descend and unable 
to mix the helium ash into hydrogen-rich material.

\begin{table}[!ht]
\label{fonts}
\caption{Supersoft X-ray count rates}
\smallskip
\begin{center}
{\tiny
\begin{tabular}{llllllllllll}
\tableline
\noalign{\smallskip}

Time &Count &Time& Count & Time& Count & Time &Count& Time &Count& Time &Count  \\

 & rate& &  rate& & rate&  & rate&  & rate & & rate \\

[day]  & [s$^{-1}$]  &  [day]  & [s$^{-1}$] &[day]  & [s$^{-1}$]&[day]  & [s$^{-1}$]&[day]  & [s$^{-1}$]& [day] & [s$^{-1}$]\\
\noalign{\smallskip}
\tableline
\noalign{\smallskip}
  3.53 &   -0.20 &  36.74 &    1.64 & 40.52 &    1.56 & 47.09 &    2.07 & 52.67 &    2.10 & 64.59 &    2.00\\
 11.33 &   -0.69 &  36.77 &    1.59 & 40.54 &    1.47 & 47.16 &    2.10 & 52.74 &    2.08 & 66.12 &    2.00\\
 11.40 &   -0.95 &  36.79 &    1.54 & 40.59 &    1.52 & 47.37 &    2.12 & 52.81 &    2.06 & 66.19 &    2.00\\
 11.47 &   -0.78 &  36.81 &    1.47 & 40.61 &    1.53 & 47.42 &    2.08 & 52.86 &    2.09 & 66.26 &    1.98\\
 13.94 &   -0.93 &  36.84 &    1.50 & 40.68 &    1.72 & 47.44 &    2.14 & 52.93 &    1.99 & 67.12 &    1.99\\
 15.96 &   -0.86 &  36.86 &    1.49 & 41.07 &    1.12 & 47.49 &    2.10 & 52.97 &    2.09 & 67.14 &    1.99\\
 18.53 &   -1.02 &  36.91 &    1.43 & 41.12 &    1.62 & 47.51 &    2.12 & 52.99 &    2.11 & 67.18 &    1.99\\
 18.57 &   -0.80 &  36.93 &    1.46 & 41.14 &    1.70 & 47.83 &    2.10 & 53.04 &    2.11 & 67.21 &    1.95\\
 26.33 &   -0.51 &  36.98 &    1.22 & 41.19 &    1.74 & 47.90 &    2.12 & 53.06 &    2.12 & 67.25 &    1.99\\
 26.35 &   -0.51 &  37.00 &    1.20 & 41.21 &    1.74 & 48.02 &    2.09 & 53.18 &    2.10 & 67.28 &    1.94\\
 29.34 &    1.07 &  37.05 &    1.11 & 41.26 &    2.00 & 48.04 &    2.09 & 53.20 &    2.11 & 67.32 &    1.97\\
 29.36 &    1.06 &  37.12 &    1.37 & 41.28 &    2.02 & 48.09 &    2.09 & 53.25 &    2.09 & 67.35 &    1.86\\
 30.22 &    0.48 &  37.14 &    1.39 & 41.33 &    2.04 & 48.11 &    2.12 & 53.27 &    2.09 & 67.51 &    2.02\\
 30.24 &    0.59 &  37.18 &    1.39 & 41.35 &    2.04 & 48.16 &    2.08 & 53.32 &    2.12 & 67.53 &    2.02\\
 32.35 &    1.21 &  37.21 &    1.41 & 41.40 &    2.04 & 48.18 &    2.13 & 53.34 &    2.14 & 68.67 &    1.99\\
 32.37 &    1.19 &  37.25 &    1.58 & 41.42 &    2.07 & 48.23 &    2.14 & 53.39 &    1.79 & 68.74 &    1.96\\
 33.16 &    1.78 &  37.32 &    1.52 & 41.47 &    1.97 & 48.30 &    2.12 & 53.41 &    1.80 & 69.66 &    1.96\\
 33.25 &    1.86 &  37.39 &    1.64 & 42.46 &    2.09 & 48.37 &    2.11 & 53.46 &    2.10 & 69.68 &    1.96\\
 33.30 &    1.95 &  37.42 &    1.56 & 42.53 &    2.03 & 48.43 &    2.10 & 53.48 &    2.10 & 69.73 &    1.97\\
 33.32 &    1.88 &  37.46 &    1.81 & 43.13 &    1.90 & 48.50 &    2.07 & 53.53 &    2.09 & 69.75 &    1.95\\
 33.37 &    1.88 &  37.49 &    1.79 & 43.20 &    1.87 & 48.57 &    2.13 & 53.55 &    2.06 & 70.59 &    1.95\\
 33.39 &    1.94 &  37.53 &    1.81 & 43.27 &    1.98 & 48.90 &    2.10 & 53.60 &    2.04 & 70.61 &    1.97\\
 33.43 &    1.94 &  37.60 &    1.84 & 43.34 &    1.97 & 49.04 &    2.10 & 53.62 &    1.97 & 70.66 &    1.96\\
 33.46 &    1.94 &  37.67 &    1.92 & 43.41 &    1.92 & 49.11 &    2.09 & 53.67 &    2.05 & 70.68 &    1.96\\
 33.50 &    1.85 &  37.74 &    1.85 & 43.48 &    1.78 & 49.18 &    2.09 & 53.69 &    1.99 & 72.35 &    1.90\\
 33.53 &    1.88 &  37.79 &    1.88 & 43.55 &    1.94 & 49.22 &    2.08 & 53.74 &    1.88 & 72.37 &    1.87\\
 33.57 &    1.92 &  37.81 &    1.50 & 43.94 &    1.69 & 49.24 &    2.10 & 53.78 &    2.03 & 72.42 &    1.89\\
 33.60 &    1.86 &  38.13 &    2.10 & 44.01 &    1.91 & 49.29 &    2.05 & 53.80 &    2.01 & 72.49 &    1.88\\
 33.64 &    1.18 &  38.32 &    2.13 & 44.08 &    1.91 & 49.31 &    2.07 & 53.85 &    1.79 & 73.30 &    1.91\\
 33.67 &    0.96 &  38.34 &    2.11 & 44.15 &    1.81 & 49.36 &    2.07 & 53.87 &    2.05 & 73.34 &    1.89\\
 33.71 &    0.37 &  38.39 &    2.14 & 44.20 &    1.67 & 49.38 &    2.08 & 53.92 &    2.09 & 73.37 &    1.87\\
 33.74 &    0.34 &  38.41 &    2.11 & 44.22 &    1.76 & 49.43 &    2.06 & 53.94 &    2.03 & 73.41 &    1.89\\
 33.78 &    0.27 &  38.46 &    2.10 & 44.27 &    1.91 & 49.45 &    2.04 & 53.99 &    2.10 & 73.43 &    1.86\\
 33.90 &    1.34 &  38.48 &    2.07 & 44.29 &    1.90 & 49.50 &    2.06 & 54.01 &    2.11 & 74.22 &    1.82\\
 33.97 &    1.66 &  38.53 &    1.97 & 44.34 &    1.43 & 49.52 &    2.06 & 54.06 &    2.08 & 74.29 &    1.79\\
 34.04 &    1.94 &  38.60 &    1.92 & 44.36 &    1.73 & 49.82 &    2.10 & 54.13 &    1.91 & 74.36 &    1.80\\
 34.11 &    1.98 &  38.67 &    2.02 & 44.41 &    1.93 & 49.85 &    2.12 & 54.18 &    2.08 & 74.43 &    1.87\\
 34.18 &    1.99 &  38.74 &    2.05 & 44.43 &    1.97 & 50.03 &    2.12 & 54.20 &    2.12 & 74.48 &    1.87\\
 34.20 &    2.01 &  38.78 &    2.02 & 44.48 &    1.95 & 50.05 &    2.13 & 54.24 &    1.78 & 74.50 &    1.88\\
 34.24 &    1.68 &  38.85 &    1.54 & 44.50 &    1.97 & 50.10 &    2.09 & 54.27 &    1.78 & 75.17 &    1.87\\
 34.27 &    1.56 &  38.92 &    0.84 & 44.55 &    1.99 & 50.17 &    2.08 & 54.31 &    2.09 & 75.22 &    1.86\\
 34.31 &    1.28 &  38.99 &    0.92 & 45.15 &    1.92 & 50.24 &    2.07 & 54.34 &    2.10 & 75.24 &    1.85\\
 34.34 &    1.45 &  39.06 &    1.33 & 45.17 &    1.90 & 50.31 &    2.04 & 54.41 &    2.10 & 77.23 &    1.81\\
 34.36 &    1.57 &  39.13 &    1.70 & 45.22 &    2.09 & 50.38 &    2.09 & 54.45 &    2.11 & 77.25 &    1.79\\
 34.38 &    1.69 &  39.34 &    1.55 & 45.29 &    2.11 & 50.52 &    2.16 & 54.48 &    2.12 & 77.30 &    1.81\\
 34.43 &    1.90 &  39.41 &    1.01 & 45.36 &    1.98 & 51.51 &    2.12 & 57.60 &    2.03 & 77.37 &    1.81\\
 34.45 &    1.95 &  39.48 &    0.88 & 45.43 &    2.12 & 51.54 &    2.11 & 57.62 &    2.05 & 79.15 &    1.74\\
 34.50 &    2.07 &  39.52 &    1.19 & 45.47 &    1.90 & 51.84 &    2.14 & 58.62 &    2.03 & 79.18 &    1.74\\
 34.52 &    2.04 &  39.55 &    1.20 & 45.49 &    1.85 & 51.91 &    2.11 & 58.64 &    1.99 & 80.52 &    1.67\\
 35.05 &    1.86 &  39.59 &    1.24 & 45.56 &    2.14 & 51.93 &    2.12 & 59.15 &    2.11 & 80.56 &    1.68\\
 36.26 &    1.51 &  39.62 &    1.27 & 45.82 &    2.09 & 51.98 &    2.12 & 59.87 &    2.07 & 80.59 &    1.68\\
 36.30 &    1.51 &  39.66 &    1.16 & 45.89 &    2.00 & 52.05 &    2.11 & 59.89 &    2.08 & 80.63 &    1.67\\
 36.33 &    1.52 &  39.68 &    1.08 & 46.03 &    2.09 & 52.12 &    2.12 & 61.63 &    2.06 & 80.66 &    1.68\\
 36.37 &    1.54 &  39.75 &    1.41 & 46.07 &    2.02 & 52.18 &    2.10 & 61.70 &    2.05 & 85.15 &    1.36\\
 36.40 &    1.53 &  39.82 &    1.33 & 46.10 &    2.02 & 52.25 &    2.12 & 62.16 &    2.01 & 85.19 &    1.35\\
 36.44 &    1.59 &  39.85 &    1.58 & 46.14 &    2.09 & 52.30 &    2.09 & 62.18 &    2.00 & 85.22 &    1.35\\
 36.47 &    1.62 &  39.87 &    1.61 & 46.17 &    2.09 & 52.32 &    2.12 & 62.23 &    2.05 & 87.14 &    1.17\\
 36.51 &    1.50 &  39.99 &    1.56 & 46.21 &    2.08 & 52.37 &    2.12 & 62.25 &    2.04 & 87.16 &    1.18\\
 36.54 &    1.43 &  40.05 &    1.66 & 46.24 &    2.08 & 52.39 &    2.14 & 62.58 &    2.08 & 87.21 &    1.17\\
 36.58 &    1.38 &  40.12 &    1.55 & 46.28 &    2.08 & 52.44 &    2.12 & 62.65 &    2.07 & 87.28 &    1.17\\
 36.61 &    1.20 &  40.19 &    1.44 & 46.30 &    2.09 & 52.46 &    2.13 & 63.18 &    2.02 & 87.35 &    1.15\\
 36.65 &    1.47 &  40.26 &    1.11 & 46.35 &    1.97 & 52.53 &    2.08 & 63.25 &    2.04 & 91.03 &    0.70\\
 36.68 &    1.41 &  40.33 &    0.87 & 46.37 &    1.97 & 52.58 &    2.10 & 64.18 &    2.04 & 91.10 &    0.74\\
 36.70 &    1.70 &  40.40 &    1.01 & 46.42 &    2.09 & 52.60 &    2.09 & 64.24 &    1.83 & 93.57 &    0.43\\
 36.72 &    1.65 &  40.47 &    1.53 & 46.49 &    2.06 & 52.65 &    2.10 & 64.52 &    1.98 & 93.90 &    0.36\\
\noalign{\smallskip}
\tableline\
\end{tabular}
}
\end{center}
\end{table}

\normalsize
\begin{table}[!ht]
\label{fonts2}
\caption{Evolutional data of the outburst model}
\smallskip
\begin{center}
{\tiny
\begin{tabular}{lllllll}
\tableline
\noalign{\smallskip}
Time &$\log T_{\rm ph}$&$\log R_{\rm ph}$& $\log$ (mass loss rate)&$\log V$  & $\log L$ &  $\log g$ \\

[day]& [K]&[cm] &[$M_{\odot}~$yr$^{-1}$] & [cm~s$^{-1}$]&  [erg~s$^{-1}$] &[cm~s$^{-2}$]\\
\noalign{\smallskip}
\tableline
\noalign{\smallskip}
        9.8 &  4.86 & 11.04 &   -4.754 &  7.990 &  38.38 &  4.181 \\
       22.1 &  5.09 & 10.60 &   -5.287 &  8.087 &  38.40 &  5.060 \\
       27.3 &  5.16 & 10.46 &   -5.507 &  8.096 &  38.41 &  5.342 \\
       32.6 &  5.29 & 10.20 &   -5.778 &  7.971 &  38.42 &  5.846 \\
       34.8 &  5.37 & 10.05 &   -5.952 &  7.815 &  38.43 &  6.151 \\
       38.2 &  5.50 &  9.79 &   -6.490 &  7.405 &  38.43 &  6.679 \\
       38.9 &  5.55 &  9.69 &   -6.820 &  7.154 &  38.44 &  6.873 \\
       39.1 &  5.56 &  9.67 &   -6.921 &  7.067 &  38.44 &  6.918 \\
       39.5 &  5.63 &  9.53 &    0.000 &  0.000 &  38.44 &  7.191 \\
       40.5 &  5.82 &  9.15 &    0.000 &  0.000 &  38.44 &  7.948 \\
       44.9 &  5.95 &  8.89 &    0.000 &  0.000 &  38.44 &  8.471 \\
       50.7 &  6.03 &  8.73 &    0.000 &  0.000 &  38.43 &  8.796 \\
       60.6 &  6.10 &  8.58 &    0.000 &  0.000 &  38.41 &  9.093 \\
       71.7 &  6.13 &  8.51 &    0.000 &  0.000 &  38.39 &  9.239 \\
       78.1 &  6.14 &  8.44 &    0.000 &  0.000 &  38.30 &  9.368 \\
       80.8 &  6.13 &  8.43 &    0.000 &  0.000 &  38.24 &  9.399 \\
       85.0 &  6.12 &  8.41 &    0.000 &  0.000 &  38.16 &  9.425 \\
       89.5 &  6.09 &  8.40 &    0.000 &  0.000 &  38.02 &  9.445 \\
       94.4 &  6.04 &  8.40 &    0.000 &  0.000 &  37.82 &  9.460 \\
       97.5 &  6.01 &  8.39 &    0.000 &  0.000 &  37.70 &  9.466 \\
      100.3 &  5.99 &  8.39 &    0.000 &  0.000 &  37.58 &  9.469 \\
      116.4 &  5.85 &  8.39 &    0.000 &  0.000 &  37.03 &  9.480 \\
\noalign{\smallskip}
\tableline\
\end{tabular}
}
\end{center}
\end{table}

%%% MAIN BODY OF TEXT GOES HERE. CONSULT "INSTRUCTIONS FOR AUTHORS USING
%%% LATEX2E MARKUP", SECTIONS 2.3-2.6 FOR HELP WITH EQUATIONS, FIGURES,
%%% AND TABLES.

%%% Fig.1
%%%\clearpage
\begin{figure}[!ht]
\plottwo{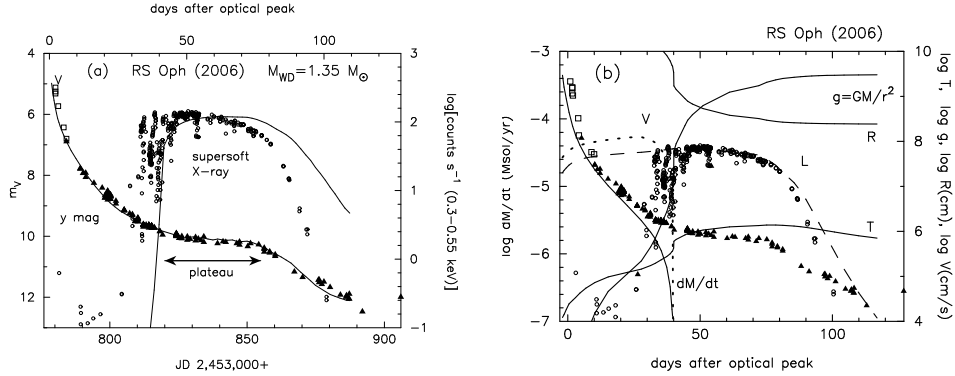}{kato_fig1b.epsi}
\caption{(a) The observational light curves of optical ($y$ band)
and supersoft X-ray (0.3-0.55 keV) as well as our best fit model
of $1.35 M_{\odot}$ WD. 
(b) The evolution of physical values in our best fit model.
The wind mass-loss rate, wind velocity at the photosphere, photospheric
temperature, photospheric radius, total luminosity, and the surface
gravity $g \equiv GM/r^2$.
The scale of the total luminosity $\log L$ is 37.0 (bottom) and 39.25 (top).
All the values are plotted in the logarithmic scale. The X-ray count rates
and evolution data are tabulated in Tables 1 and 2. The origin of time
is set to be JD 2453779.
}
\label{bestfitmodel}
\end{figure}

%\acknowledgements %%% Text of acknowledgements runs on after this command.

%%% THE BIBLIOGRAPHY
%%%
%%% CONSULT SECTION 3 OF "INSTRUCTIONS FOR AUTHORS" FOR HOW TO USE NATBIB.
%%% AUTHORS ARE ENCOURAGED TO USE EITHER THE "THEBIBLIOGRAPY" ENVIRONMENT
%%% BY UNCOMMENTING (DELETING THE "%" SYMBOL) THE COMMANDS BELOW, OR BY
%%% USING THE BIBTEX ENVIRONMENT. TO FIND OUT WHICH IS APPLICABLE TO YOUR
%%% CONTRIBUTION, CONSULT THE VOLUME EDITORS FOR YOUR PROCEEDINGS.
%%%

%reference

\end{document}